\documentclass{article}

\PassOptionsToPackage{numbers, compress}{natbib}

\usepackage{graphicx} 
\usepackage{placeins}
\usepackage[preprint]{neurips_2026}
\usepackage{float} 
\usepackage[utf8]{inputenc} % allow utf-8 input
\usepackage[T1]{fontenc}    % use 8-bit T1 fonts
\usepackage{hyperref}       % hyperlinks
\usepackage{url}            % simple URL typesetting
\usepackage{booktabs}       % professional-quality tables
\usepackage{makecell}    % For easy multiline table cells
\usepackage{amsfonts}       % blackboard math symbols
\usepackage{nicefrac}       % compact symbols for 1/2, etc.
\usepackage{microtype}      % microtypography
\usepackage{xcolor}         % colors
\usepackage{adjustbox}      % adjust box around figures
\usepackage{enumitem}       % helps with bullet points

\title{The Differentiable Auditory Loop (DAL): 
  An ML Framework for Hyper-Personalized Hearing Aids 
}

\author{%
  Alejandro Ballesta Rosen$^{1}$ \\
  \texttt{alejandrobr@google.com}
  \And
  Jason Mikiel-Hunter$^{2}$ \\
  \texttt{jason.mikiel-hunter@mq.edu.au}
  \And
  Julian Maclaren$^{1,2}$ \\
  \texttt{jmaclaren@google.com}
  \And
  Jack Collins$^{1}$ \\
  \texttt{jackcollins@google.com}
  \And
  Richard F. Lyon$^{1}$ \\
  \texttt{dicklyon@google.com}
  \And
  Simon Carlile$^{1,2}$ \\
  \texttt{scarlile@google.com}
  \\[1.5em]
  $^{1}$Google Research Australia \\
  $^{2}$Macquarie University
}

\begin{document}

\maketitle

\begin{abstract}
Conventional hearing aids rely on fixed, frequency-dependent amplification and compression to manage reduced sensitivity, which often fails to provide sufficient listening support in complex environments, such as situations with multiple speakers (the ``cocktail party'' problem). To more comprehensively address the underlying encoding dysfunctions of hearing loss, we introduce the Differentiable Auditory Loop (DAL), a new open-source framework for personalized hearing aid design and fitting. Our first implementation of DAL incorporates CARFAC, a differentiable model of human cochlear function, which we ported to JAX, to optimize a deep neural network to match impaired auditory neural activity patterns with a normal-hearing reference. To build a hearing aid with the fine-grained spectro-temporal signal processing required, we adopt SEANet, a waveform-to-waveform fully convolutional UNet generator. We fine-tune the network by comparing the outputs of a CARFAC model fitted to normal hearing with that of a CARFAC model fitted to match each subject's individual hearing impairment. The comparison is done using loss functions derived from the respective CARFAC neural activity pattern (NAP) outputs and stabilized auditory images (SAIs), the latter providing a 2D representation that captures phase-insensitive temporal structure in the auditory nerve output. Through gradient descent, the SEANet model learns to both denoise the input and compensate for the hearing loss modelled by the impaired CARFAC model. Across neural-representation and signal-fidelity metrics, the DAL-optimized SEANet model outperformed the tested master hearing aid (MHA) baselines. The DAL framework provides a practical path toward model-based, machine-learning-driven personalization of hearing aid signal processing. Next steps include hardware deployment to enable real-world clinical testing.
\end{abstract}

\section{Introduction}

Nearly 50 years ago, Reiner Plomp characterized hearing impairment in terms of two different processes \cite{Plomp1978-fd}. First, reduced sensitivity which is traditionally managed by linear amplification followed by compression above some input level. Second, reduction in perceptual fidelity at sound levels above threshold (suprathreshold) that interferes with the segregation and comprehension of sounds such as speech in noisy backgrounds. Plomp and others have argued that encoding dysfunctions in the inner ear underlies both of these processes \cite{Plomp1978-fd}; however, there are a wide range of inner ear processes that can fail, resulting in hearing impairment. Examples include inner and outer hair cell damage and synaptic dysfunction due to loud sound exposure \cite{Wagner2019-wo}. As a result, reduced sensitivity itself is a salient symptom of hearing problems and measured using the standard pure tone audiogram which, however, sheds no light on the underlying dysfunction. Furthermore, the linear digital signal processing (DSP) used to deal with the symptom fails to engage with the nature of the dysfunctions.

Modern commercial hearing aids continue to rely on prescriptive digital filter adjustment algorithms, such as the NAL-NL2 multi-band fitting formulation \cite{Keidser2011-nm}. The traditional paradigm adjusts a series of discrete, frequency-dependent linear amplification channels combined with a fast-acting dynamic range compressor \cite{Kates2008-zm}. This approach operates entirely symptomatically so that the fitting relies on pure-tone thresholds measured audiometrically to compensate for hearing loss in quiet using within channel gain and compression, ignoring other potential hearing dysfunctions (e.g. coding distortions). Some current hearing aids use a denoising ML model at the front end but this does not address the encoding distortions at suprathreshold levels that affect the understanding of speech in noise. In contrast to the constrained simple linear filters and static directional beamforming of legacy devices, the DAL framework reported here provides the potential to execute context-aware signal transformation that attempts to reassert the normally hearing pattern of neural signals in the auditory nerve.

\begin{figure}[htbp]
  \centering
    \includegraphics[width=0.75\linewidth]{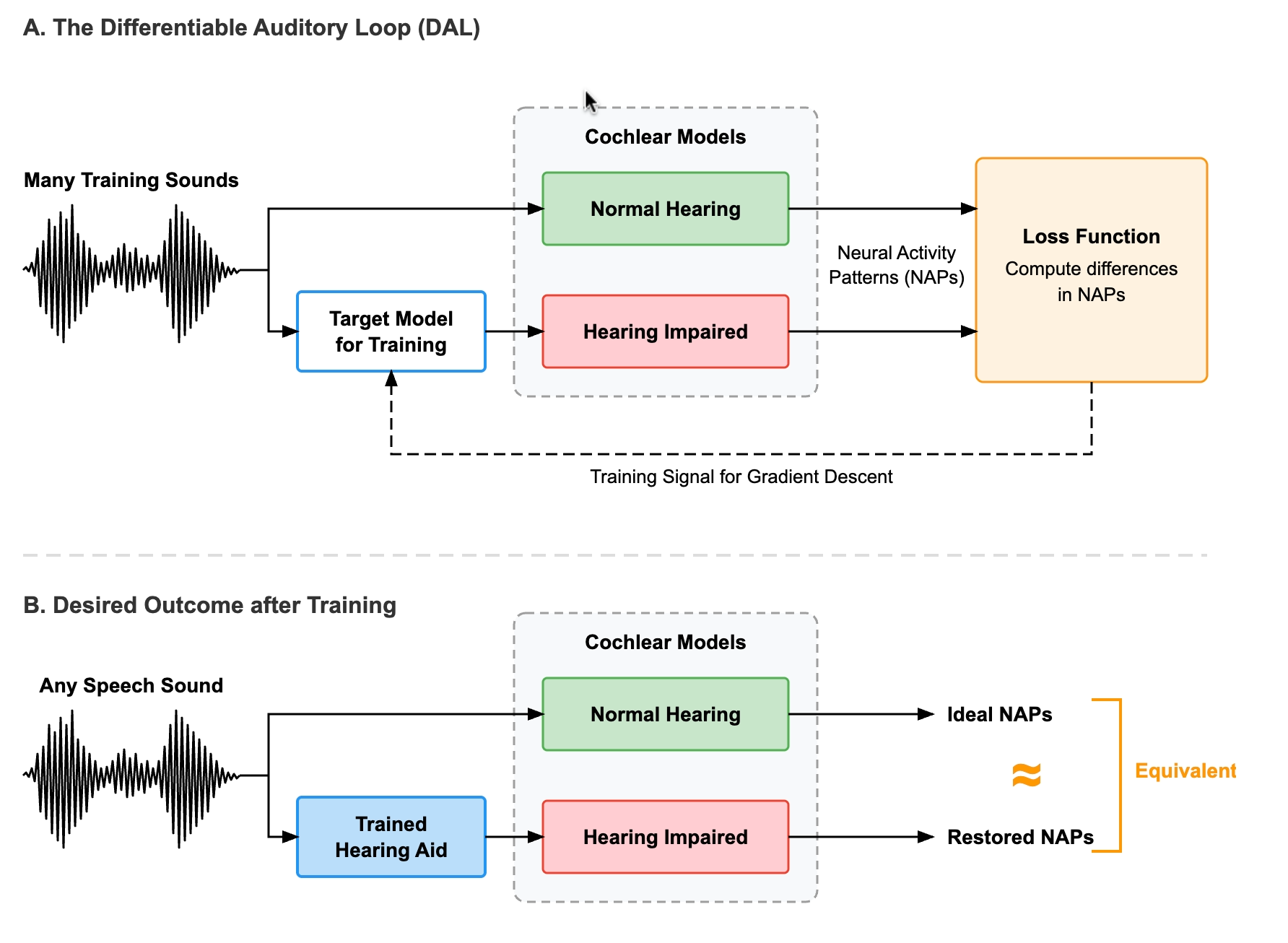}%
  \caption{A) Basic architecture of the DAL framework using computational models of the inner ear and loss functions derived from the differences in the neural activity patterns from each model. B) The goal once training is complete. The combination of the trained hearing aid and the hearing impaired model should produce neural activity patterns that are as close as possible to those from a subject who can hear normally.}
\end{figure}

The heart of the framework is an efficient and sophisticated model of hearing function of the inner ear. We chose a biomimetic computational model of the human cochlea, the Cascade of Asymmetric Resonators with Fast-Acting Compression (CARFAC) \cite{Lyon2018-pr,Lyon2024-ko}, as it explicitly models many of the major structures and processes involved in transducing the mechanical energy in a sound wave to the neural activity patterns in the auditory nerve. We extended the model to also incorporate the different synaptic connections between the inner hair cells (coding sensors) and the auditory nerve and included a middle ear transfer function. The model has been ported to JAX/FLAX and used to model both normal-hearing and impaired-hearing activity patterns on the auditory nerve (Fig. 1). This fully differentiable platform enables a sophisticated neuro-computational correction of the input to the hearing impaired ear to reassert or, at least approximate, the normally occurring neural signals in the auditory nerve to the brain. The DAL framework was developed alongside our active clinical trials involving novel hearing tests. Data from these trials will ultimately allow for fitting of personalized CARFAC models and, via use of the DAL framework, a personalized hearing aid. 

Our main contributions are:
\begin{itemize}[label={\Large\textbullet}, itemsep=0.25em, topsep=0.1pt]
    \item DAL framework proof-of-concept that can address hearing loss by targeting specific underlying dysfunctions in the inner ear through computational modelling;
    \item Porting the latest CARFAC ``v3'' version to JAX, and integrating it into the DAL framework as a biologically inspired model, as opposed to the perceptual models used by traditional hearing aid fitting; 
    \item Applying perceptually motivated loss functions to target recovery of auditory information via our ML hearing aid. These include losses on CARFAC neural activity pattern (NAP) outputs and losses based on stabilized auditory images (SAIs), an analysis that approximates the role of more central auditory brain centres, allowing us to test both pointwise neural-alignment and feature-based minimization objectives during training.
    \item Adapting SEANet, a waveform-to-waveform low-latency CNN, as a basis for ML-based signal processing, facilitating integration into resource-constrained hearing-aid platforms;
    \item Training a low-resource machine-learning hearing aid using the DAL framework, and comparing the results to conventional processing (master hearing aid implementation); 
    \item Open-source framework: This work is part of an open-source project and all other code is either already publicly available in our project GitHub or will be made open-source on publication. The major components used in our framework (CARFAC and SEANet) are already publicly available.
\end{itemize}

\section{Related Work}

Conventional hearing aids use band-limited channel processing and some fixed directional microphone beamforming. Signal processing relies on a small number (8-22) of frequency channels and dual-stage automatic gain control (AGC) mechanisms with fast-acting attack and release compression above some threshold. One advantage of this DSP approach is low latency (< 10~ms), and we wish to preserve that. As these systems divide sound into fixed frequency bins and amplify/compress bands independently there is no deeper context regarding phonetic structure of speech input. Consequently, they fail to resolve background interference. When multiple dynamic noise and signal sources share overlapping spectral channels they are processed equally.

Over the last few years there has been increasing interest in the application of ML to hearing enhancement. Much of this has been focussed on denoising to increase the SNR of the target talker. This functionality has recently been demonstrated in commercial hearing aids \cite{Fitzgerald2025-ow}. More recently, biologically inspired modelling of hearing impairment has also been used to examine ways in which DNN processing can compensate for OHC damage and synaptopathy \cite{Leer2024-hv,Wouters2026-wr}. The work reported here differentiates from current efforts in (\textit{i}) focusing on a trainable biomimetic model of cochlear function that will enable many different forms of hearing impairment to be explored and (\textit{ii}) using an existing open-source lightweight low-latency noise-reduction network as the base model for an ML-based hearing aid, thus enabling a more rapid adoption of the outcomes in resource-constrained hearing-aid platforms. 

\section{Proposed Method}

\subsection{Differentiable Auditory Loop Architecture}

A core innovation of this work is the shift from traditional prescriptive filter tuning to an end-to-end differentiable training loop, suitable for ML. Rather than targeting perceptually modelled frequency responses, we employ the CARFAC model of the cochlea operating at suprathreshold sound levels as the central optimization surface to train a machine-learning hearing aid that better matches biology.

DAL (see Fig. 1A) comprises two distinct pathways:

\textbf{The Healthy Reference Path (green path):} A clean (or less noisy) speech waveform is passed through a CARFAC cochlear model of the unimpaired, normally hearing ear. This produces baseline, healthy neural activity patterns (NAPs) that represent normal auditory encoding of a denoised sound, an ``ideal'' target.

\textbf{The Impaired Hearing Path (blue path)}: The incoming speech signal is first processed by the DNN (acting as the adaptive hearing aid). The pre-compensated output is then passed through the impaired cochlear model that is explicitly configured to simulate the individual's hearing impairment. In the case reported here, hearing impairment is modelled as a reduction in outer hair cell (OHC) undamping.

By applying a loss function to the differences between the healthy target NAPs and the hearing-impaired NAPs, gradient descent explicitly forces the machine-learning target to apply dynamic spectro-temporal pre-compensation. Consequently, the network learns to alter the incoming sound so that the neural activity pattern from the impaired ear closely approximates that from the normally hearing ear (Fig. 1B).

\subsection{Component Selection: CARFAC and SEANet}

The DAL framework requires two main components: a model of cochlear function and an ML model to function as the actual hearing aid.

\textbf{The Cochlear Simulation (CARFAC):} We have extended CARFAC \cite{Lyon2024-ko} to include explicit representation of the three fibre/synapse types innervating the inner hair cell and then ported this to JAX as the differentiable inner ear model. CARFAC accurately simulates basilar membrane non-linear resonance, active OHC feedback loops, and inner hair cell synaptic adaptation, allowing us to parameterize both healthy hearing and a range of impairments.

\begin{figure}[htbp]
  \centering
  \adjustbox{trim=0cm 0.1cm 0cm 0.0cm, clip, margin=0.15cm 0.0cm 0.0cm 0cm}{%
    \includegraphics[width=0.9\linewidth]{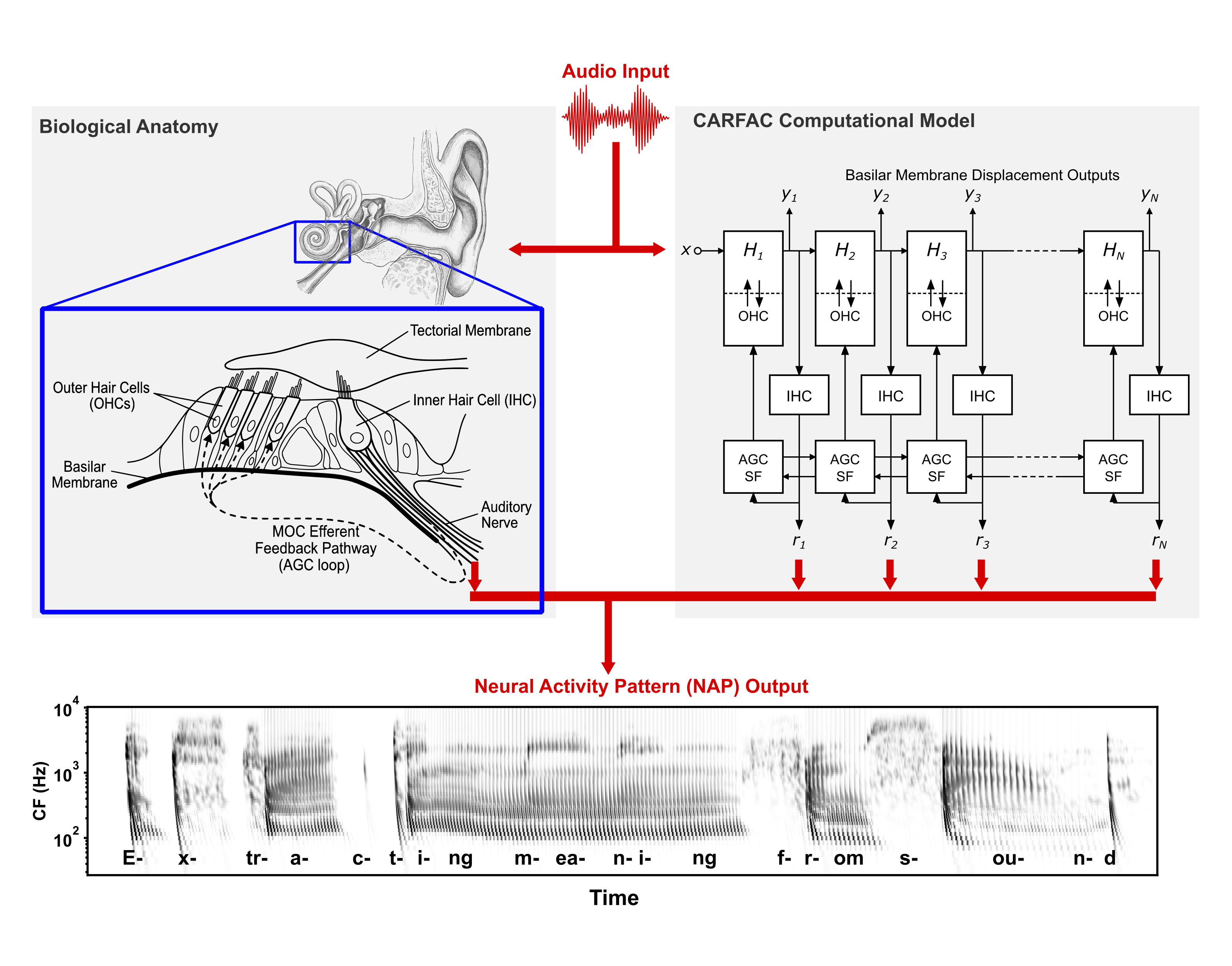}%
  }
  \caption{Structures in the cochlea (\textit{above left}) that are explicitly modelled by the CARFAC model (\textit{above right}). Both the biological anatomy and the CARFAC model take an audio input and produce a representation called a neural activity pattern (NAP) (\textit{below}). The example NAP was elicited by the spoken and recorded sentence, ``Extracting meaning from sound'', presented to normal hearing CARFAC at 80~dB SPL. CF is the characteristic frequency (in Hz), the frequency at which each channel is most sensitive.}
\end{figure}

\textbf{The Machine Learning Hearing Aid (SEANet):} As the pre-compensating machine-learning generator, we employ a low-latency causal SEANet \cite{Tagliasacchi2020-jk} (Sound EnhAncement Network) autoencoder composed of stacked multi-rate strided dilated causal convolutions mapping local temporal features. In addition to affording an integrated noise reduction capability, the model is lightweight in terms of processing and memory requirements and also has the low throughput latency required for deployment as a hearing aid. The 10~ms latency limit prevents the use of frequency-domain transformers or heavy bidirectional recurrence. The strict temporal causality guarantees no forward-looking buffer dependencies, mapping direct phase-aligned output frames within the required perceptual latency bounds.

\subsection{Loss Function -- NAPs and SAIs}
The CARFAC computational framework efficiently and accurately simulates the nonlinear sound processing performed by the human cochlea \cite{Lyon2018-pr,Saremi2016-au}. CARFAC represents each stage of its cochlear processes with digital circuits. For instance: (\textit{i}) the initial cascade of asymmetric resonators (CAR) representing basilar membrane (BM) wave propagation; (\textit{ii}) gain via active undamping by the outer hair cells, controlled by efferent AGC; and (\textit{iii}) the inner hair cell and synapse model that results in a Neural Activity Pattern (NAP) output that emulates the different classes of auditory nerve fibers. An early version of CARFAC was previously implemented in JAX to allow automatic differentiation \cite{Lyon2024-ko,Frostig_undated-ty}; in this work, we have ported the upgraded ``v3'' version of the CARFAC model (that now includes three auditory nerve fiber types) into our training loop, optimizing for parallel computation across TPU or GPU, with the ultimate goal of formulating a novel ML hearing device. Due to the transparent nature of JAX's autodifferentiation \cite{Frostig_undated-ty,UnknownUnknown-tb}, we can now target specific parameters as the potential aetiology for an individual's hearing loss and include these impaired versions of CARFAC in our final loss function for training ML hearing aids.

The frequency resolution of the CARFAC channels is 2 channels per nominal ERB, resulting in 77 channels up to 14.4~kHz when CARFAC is run at a sample rate of 32~kHz. Rather than imposing static band thresholds, the ML model acts across full-spectrum temporal fine structure, utilizing the differentiable inner ear model to determine the precise level and phase adjustments required to recover neural activity patterns that best match the normally hearing model for the same source. This enables a signal transformation based on the local context and over tens of milliseconds and many frequency bands to provide the potential to focus on the specific encoding dysfunctions underlying hearing disability rather than just increasing the sound levels within bands. 

\begin{figure}[htbp]
  \centering
  \adjustbox{trim=0cm 0.2cm 0cm 0.0cm, clip, margin=0.cm 0.cm 0.cm 0cm}{%
    \includegraphics[width=0.7\linewidth]{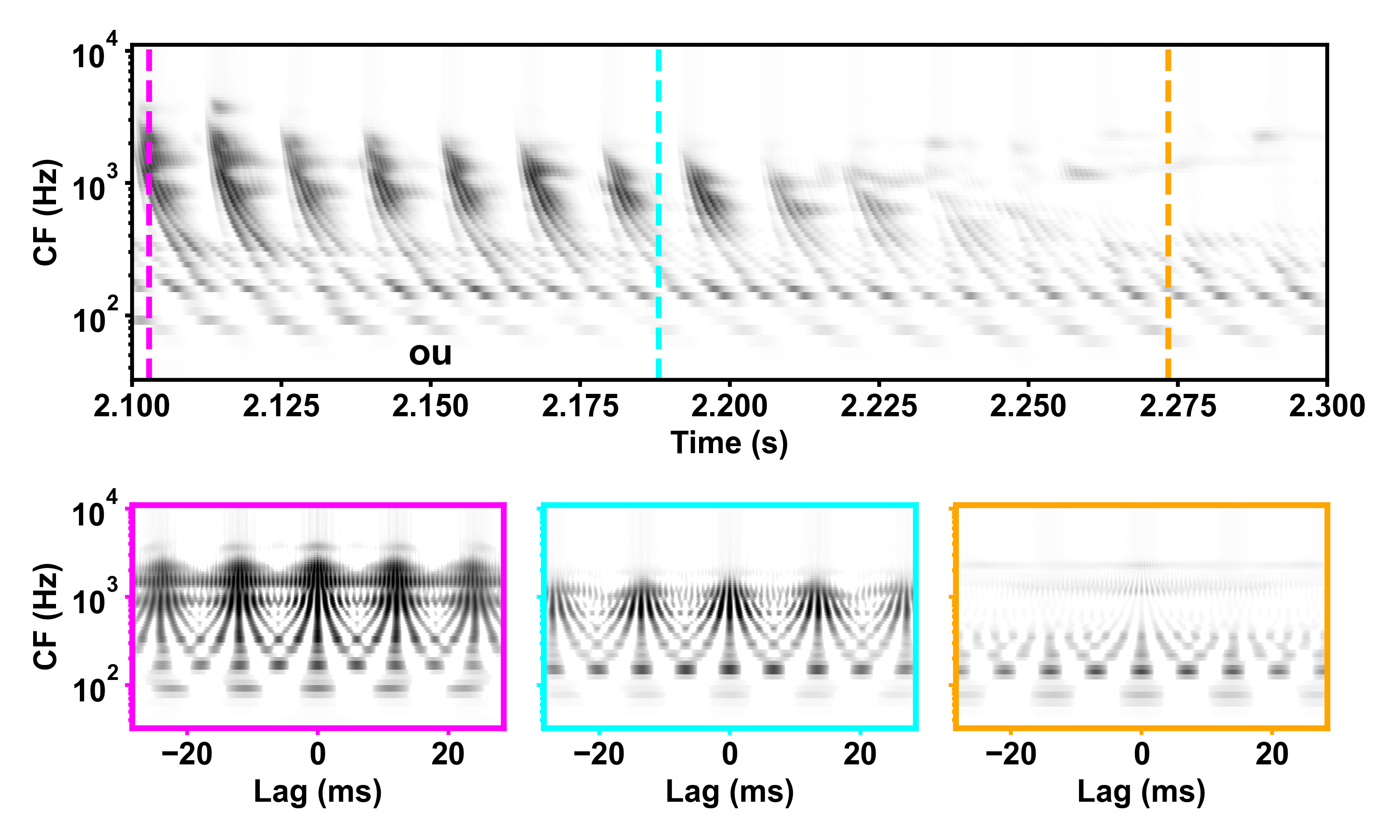}%
  }
  \caption{Normal-hearing NAPs generated for vowel ``ou'' in ``sound'' (\textit{above}). Three vertical dashed lines mark the centre of the windowed frames used to calculate SAI frames (\textit{below}).}
\end{figure}

Since reducing OHC health unavoidably alters the phase response of CAR filters, the resulting phase shifts can create misalignments when comparing the periodic NAP outputs of impaired and normal hearing CARFAC responses to the same sound. To overcome this problem and generate phase-insensitive data for our loss function, continuous ``movies'' of Stabilized Auditory Images (SAIs) were generated by applying a strided window to the NAPs and calculating their short-term autocorrelation via FFT. Each frame maps temporal lags against cochlear frequency channels on the x- and y-axes, dynamically visualizing auditory features in a manner closely aligned with human sound perception.

Determining the relevant distance between these two-dimensional SAIs is important for stable model convergence and for optimizing what matters. We have implemented the following range of loss functions driving gradient descent to find what loss function best generates a hearing aid that can recover normal NAPs and SAIs as effectively as possible:

\begin{itemize}[label={\Large---}, itemsep=0.25em, topsep=0.1pt]
    % \item \textbf{L1 Distance:} 
    \item \textbf{L1 Distance on NAP:} 
    Unnormalized mean absolute error across all pixels of cochleagram, i.e., raw NAPs. Using mean absolute error, as opposed to L2's mean squared error, helps prevent excessive bias towards high energy regions such as stimulus onsets in quiet however this distance still suffers from underestimating low-energy regions such as unvoiced consonants or fricatives. 
    
    % \item \textbf{Structural Similarity Index Measure (SSIM):} 
    \item \textbf{Structural Similarity Index Measure (SSIM) on SAI:} 
    SSIM computes local means, variances and correlations within a sliding window to calculate the difference in luminance, structure and contrast components between images \cite{Wang2004-xg, Brunet2012-mr}. The normalized luminance and contrast can often appear insensitive to differences at high levels and yet give relatively large penalties in low energy regions of stimuli \cite{Nilsson2020-xb}.
    
    % \item \textbf{Partial Normalization (PN) continuous NAP loss:} 
    \item \textbf{Partial Normalization (PN) loss on SAI:} 
    Representing a partially normalized family of distances with controllable exponents (here $\alpha$=0.5, $\beta$=0.8), which can sit between the two extremes of L1 and SSIM, PN distance retains both structural and level dependent information of the neural response.
    
    % \item \textbf{Hybrid L1/PND or SSIM loss:} 
    \item \textbf{Hybrid loss on NAP/SAI:} 
    A combination of L1 distance calculated on NAP with either PND or SSIM on SAIs was applied to test its potential benefits.
\end{itemize}

\subsection{Data Pipeline Interleaving and Sequential Batching}
Our end-to-end data pipeline splits incoming full-length audio files into contiguous temporal chunks measuring 8,192 digital samples at 32 kHz (i.e. equivalent to 0.256 s). These 8,192-sample segments are loaded into sequential batches under a global batch size of 128 using an interleaved streaming TFRecord configuration. By passing initial states forward across successive chunks via specialized carryover buffers under the Adam optimizer (beta1 = 0.5, beta2 = 0.9), the optimization loop retains unbroken temporal fine structure continuity, natively reproducing complex human forward masking responses over long multi-talker time scales.

\section{Experiments}
\subsection{Experimental Setup}
To demonstrate the feasibility of the approach, we degraded the outer hair cell (OHC) stage that is directly coupled to the `CAR' component of CARFAC and receives feedback from the automatic gain control (AGC) circuit to generate level-dependent, nonlinear tuning of the cochlear filters' gain and bandwidths. We used a depleted OHC-health parameter of 0.5 to effectively simulate a mild but significant hearing loss that exceeds 30~dB~HL between 3 and 4~kHz.

\begin{figure}[htbp]
  \centering
  \adjustbox{trim=0cm 0cm 0.1cm 0.cm, clip, margin=0.1cm 0.05cm 0.0cm 0.1cm}{%
    \includegraphics[width=0.5\linewidth]{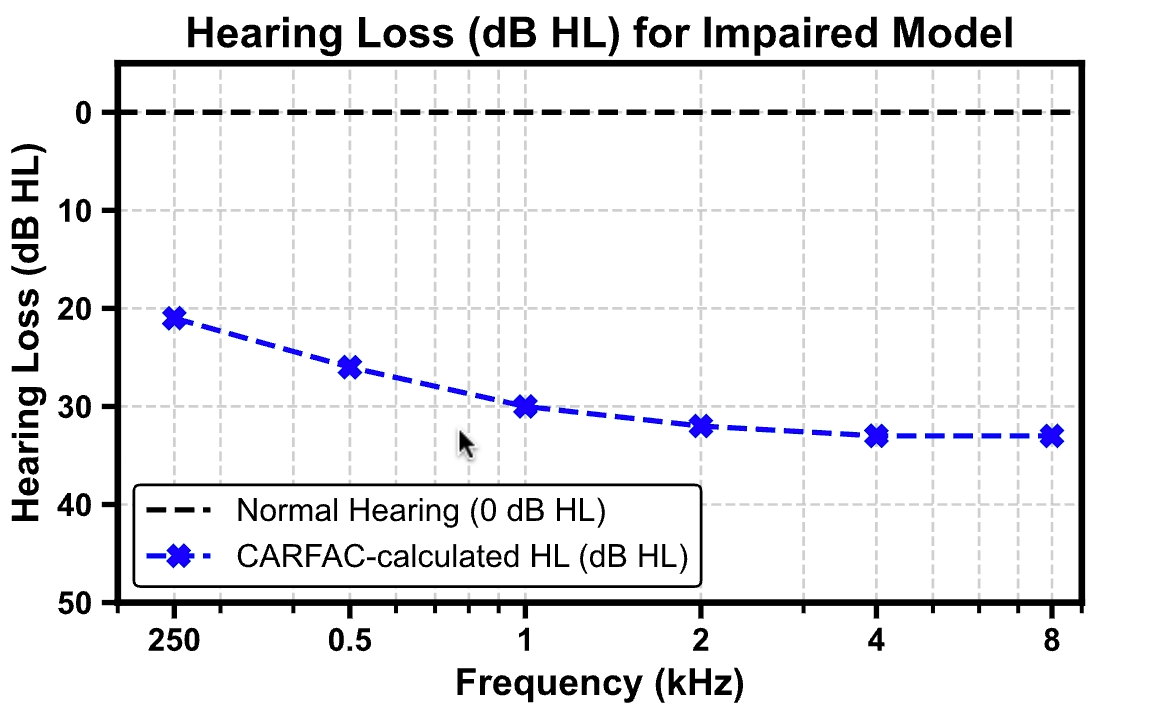}%
  }
  \caption{Audiogram showing hearing loss of impaired CARFAC model calculated as decrease in CARFAC filterbank gain.}
\end{figure}

To benchmark the enhancing and denoising capabilities of the SEANet-based hearing aid, we implemented a ``Master Hearing Aid'' (MHA) software platform in JAX to simulate conventional hearing aid amplification. By incorporating 
multi-band dynamic range compression (MBDRC), this software offers a number of modifiable parameters including target gains as well as the amount of compression (compression ratio) that activates above specific dB levels (compression thresholds) in each of 16 frequency bands.  For the purposes of this study, two different MHA configurations were utilized: 1. Gain/compression settings proposed by the NAL-NL2 formula, which prescribes hearing aid parameters according to an individual's audiogram \cite{Keidser2011-nm}; 2. Gain/compression parameters that were adjusted using the DAL training loop (clean audio only) with the MHA set as input to the impaired CARFAC model.

\subsection{Dataset}
LibriSpeech consists of English speech derived from audiobooks and sampled at 16 kHz \cite{Panayotov2015-sd}. For all experiments, we used 10,265 4-second chunks from the ``train-clean-100'' portion of the dataset. To ensure the model generalizes to unseen voices, the audio files were partitioned into training and testing sets based on speaker ID. One speaker was randomly selected and held out exclusively for the test set, while all remaining speakers were used for training. The audio files were segmented into non-overlapping chunks of 4 seconds in duration. Any remaining audio at the end of a file that was shorter than 4 seconds was padded with zeros to maintain a uniform input length of 128,000 samples at the target sample rate of 32 kHz (after upscaling from their original 16 kHz sample rate). The clean speech chunks were calibrated to a random intensity level between 50 and 80 dB SPL to simulate varying speech loudness. For each chunk, a unique segment of white noise was generated. The noise was scaled and mixed with the calibrated clean speech to achieve a target Signal-to-Noise Ratio (SNR) randomly selected between -5 dB and 10 dB. 

During training, each 4-second TFRecord example was further subdivided into contiguous 8,192-sample training chunks.

\begin{figure}[htbp]
  \centering
  \adjustbox{trim=0.3cm 0.1cm 0.3cm 0.cm, clip, margin=0.0cm 0.0cm 0.0cm 0.0cm}{%}
  \includegraphics[width=1.01\linewidth]{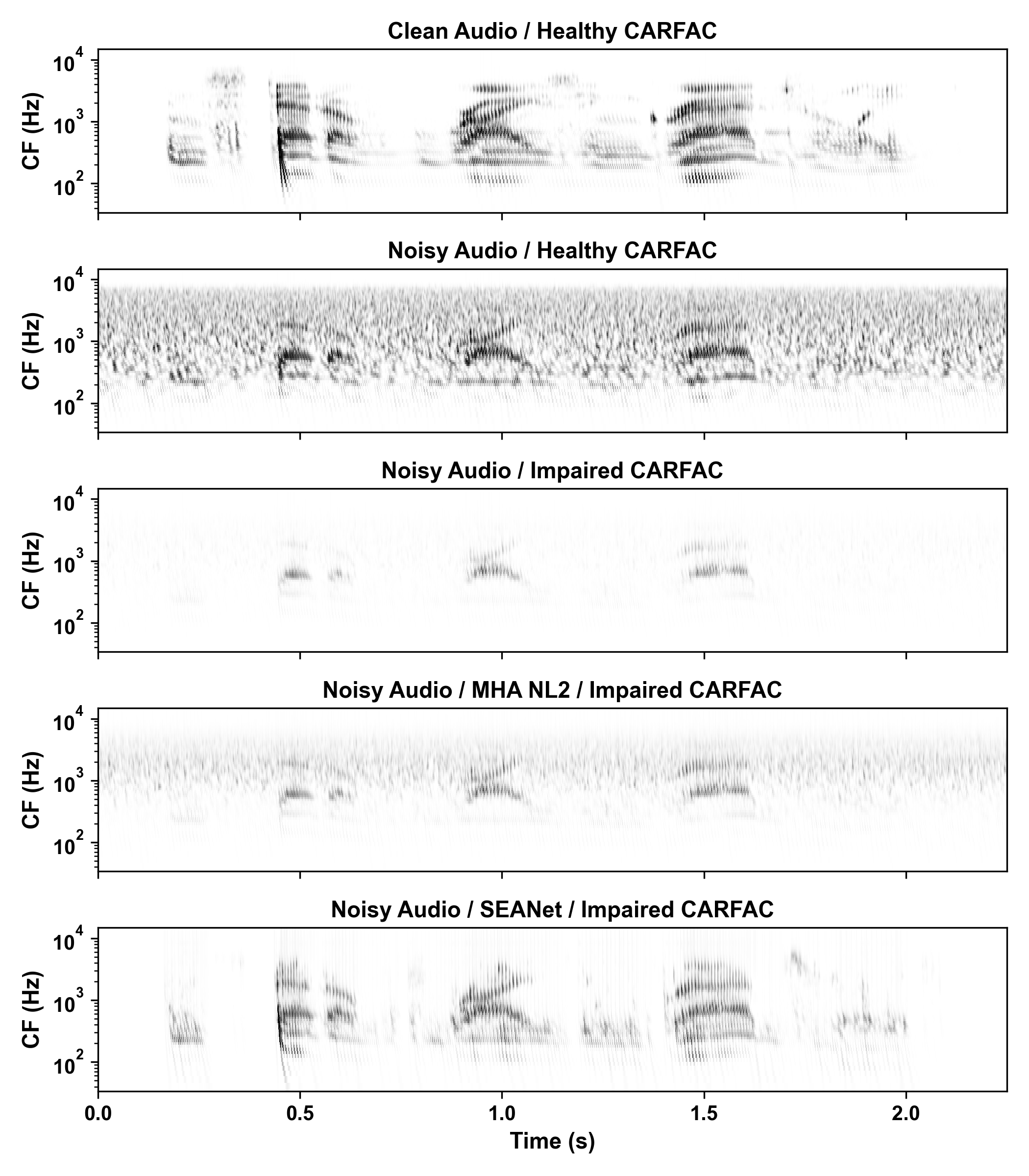}
  }
  \caption{Visual comparison of Neural Activity Patterns (NAPs). The phrase, "instead of rising gradually" was presented in quiet (\textit{Clean Audio}, level: 59.6dB SPL) or in white noise (\textit{Noisy Audio}, SNR: -2.8dB) conditions Top to bottom: (1) clean audio processed through a normal-hearing CARFAC, (2) noisy audio processed through a normal-hearing CARFAC, (3) noisy audio processed through an impaired CARFAC, (4) MHA-processed noisy audio through an impaired CARFAC, and (5) SEANet-enhanced noisy audio through an impaired CARFAC. The SEANet output suppresses background noise and more closely reconstructs the auditory structural features present in the clean normal-hearing reference.}
  \label{fig:nap-comparison}
\end{figure}

\subsection{Quantitative Results: Benchmarking of Differentiable Loss Outcomes}

To evaluate the efficacy of the proposed SEANet architectures guided by CARFAC-based loss functions, we compared their performance against both the unprocessed noisy input (Baseline) and a standard Master Hearing Aid (MHA) processing baseline (e.g., MHA NL2) (see example NAP responses in Fig. 5). The evaluation strictly measures distance and similarity against clean references across neural representation metrics---Neural Activity Pattern (NAP) and Stabilized Auditory Image (SAI). A scale-invariant signal-to-distortion ratio (SI-SDR) metric was also calculated on the NAPs.

\begin{figure}[htbp]
  \centering
    \includegraphics[width=1\linewidth]{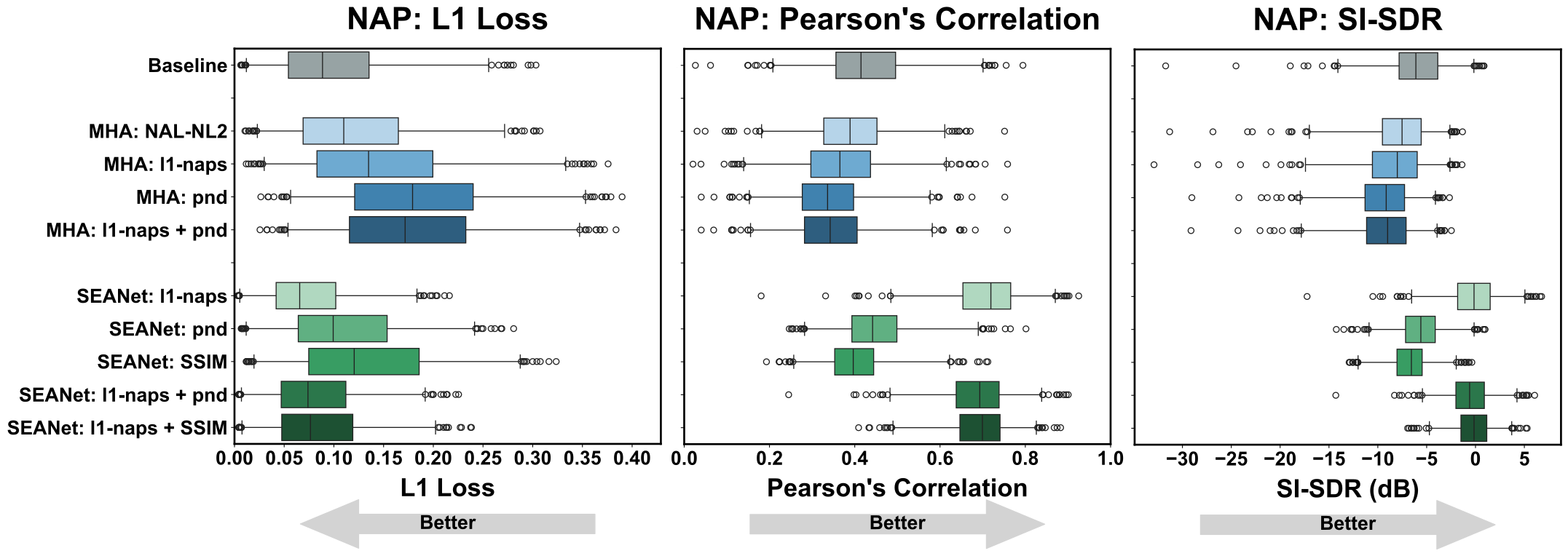}
\caption{Comparison of NAP metrics for unaided and aided audio in noise. Metrics were evaluated across \textit{n}=1000 4-second utterances. L1 loss (\textit{left}), Pearson's correlation (\textit{middle}), and SI-SDR (\textit{right}) capture the similarity between a reference NAP (clean audio, normal-hearing CARFAC) and experimental NAPs (unaided or aided noisy audio, impaired CARFAC). Conditions shown include the unaided baseline (\textit{grey}), MHA (\textit{blue}), and SEANet (\textit{green}). Grey arrows indicate improved similarity to the reference. Box plots indicate median (vertical line), interquartile range (IQR, box edges), 1st and 99th percentiles (whiskers), and outliers beyond this range (circles).} 
\end{figure}

\textcolor{black!80}{\textbf{Neural Representation Recovery}}\\[1.5pt]
Overall, SEANet models optimized with CARFAC-based losses tended to successfully reconstruct the auditory structural features of the clean speech, outperforming both the unprocessed baseline and standard MHA processing. 

In the NAP domain, standard MHA processing consistently degrades responses, resulting in higher L1 distances and lower correlations than the baseline (unprocessed noisy audio presented to impaired CARFAC) (Fig 6 and Supplemental Table C). The fact that the trained MHA is worse on these objective metrics than the untrained \texttt{mha (nl2)} implies that our training loss functions are not well aligned with these objective measures, which is not surprising given that the MHA was trained on clean audio. In contrast, SEANet models demonstrated substantial improvements in the pointwise NAP metrics, provided that the L1 loss on the NAP representation (\texttt{l1-nap}) had not been included during training.  The \texttt{seanet (l1-nap)} model's denoising capability reduces the L1 distance and improves the correlation most, rendering the output closest to the clean reference. Furthermore, any combination utilizing this loss term yields clear gains over the baseline. However, optimizing SEANet without the direct \texttt{l1-nap} penalty proved ineffective in this domain and actually degraded these NAP metrics relative to the noisy baseline. Again this can be explained by phase relationships not being constrained in training based on SAI metrics.

\begin{figure}[htbp]
  \centering
  \includegraphics[width=0.75\linewidth]{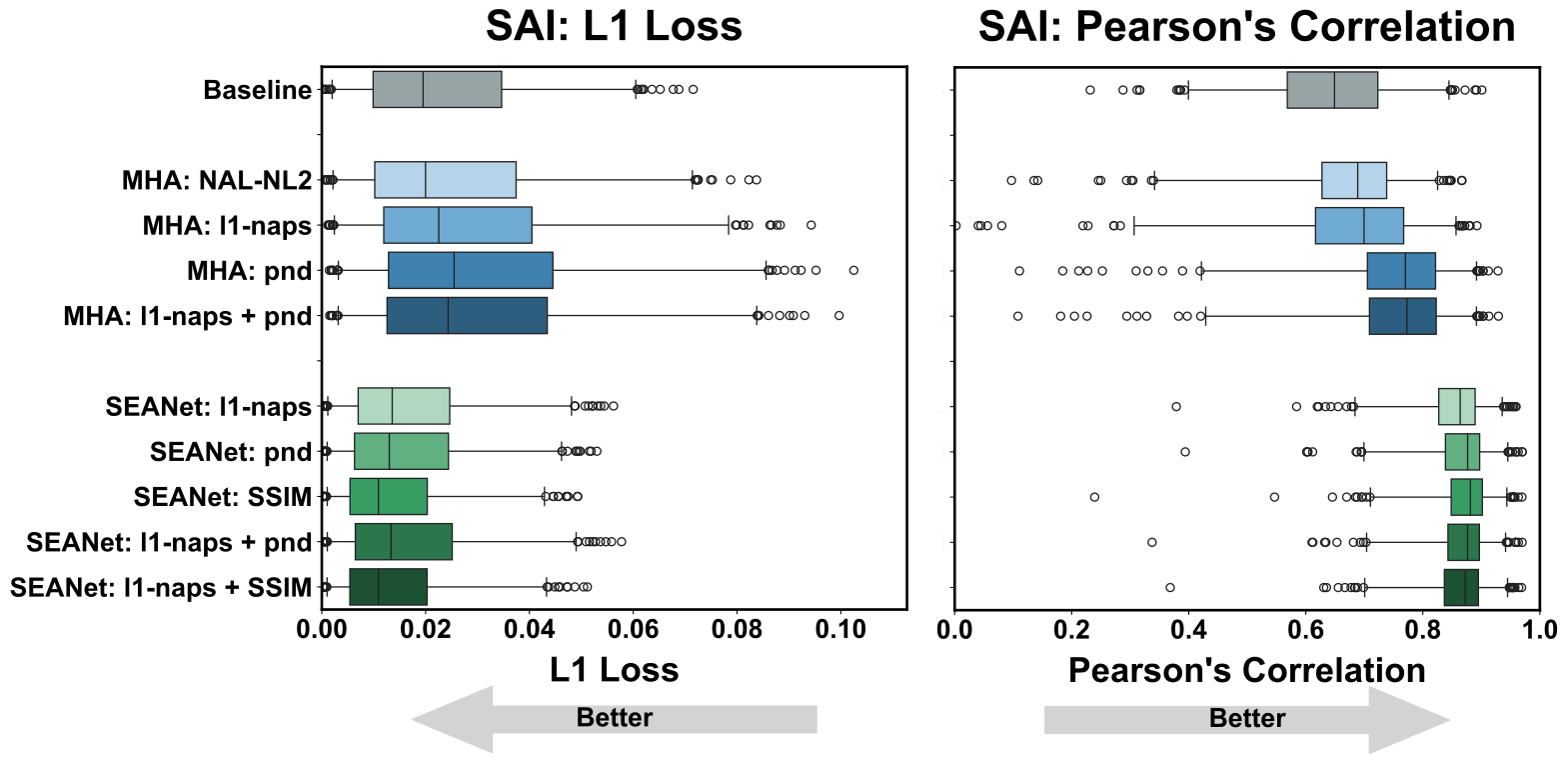}
  \caption{Comparison of SAI metrics for unaided and aided audio in noise. L1 loss (\textit{left}) and Pearson's correlation (\textit{right}) were calculated by comparing reference SAIs (clean audio processed by a normal-hearing CARFAC model) against experimental SAIs (unaided or aided audio processed by an impaired CARFAC model). Experimental conditions, model variants (colors), and box plot conventions are identical to those detailed in Fig. 6.}
\end{figure}

SEANet's improvement over the noisy baseline and MHA versions is also observed in the SAI domain (Fig. 7), where temporal acoustic structure is represented in a phase-insensitive form (Fig. 7 and Supplemental Table D). Standard MHA processing presents a mixed impact on these representations; while it generally improves the Pearson correlation (for example, \texttt{mha (nl2)}), it consistently exacerbates the L1 distance. In contrast, all SEANet configurations provide substantial improvements. Notably, the dynamics between the loss functions shift considerably in this domain compared to the NAP domain, as applying the \texttt{l1-nap} loss does not now provide an outright benefit. Instead, models optimized solely with \texttt{pnd} or \texttt{ssim} prove highly competitive on their own and the latter proves the most effective and rendering SAIs that most closely match those observed for the clean audio presented to the normal hearing CARFAC.

Computing the standard error of the mean (SEM) as the sample standard deviation divided by the square root of the number of samples gives very low SEM values due to the 1000 samples used, However, these tight spreads to do not include the expected effect of different random training runs, which is likely to dominate. So we are not yet in a position to assess the statistical significance of the differences seen in Figures 6 and 7

\textcolor{black!80}{\textbf{Fidelity of the Neural Representation}}\\[1.5pt]
The structural improvements (L1 distance and Pearson correlation) observed in these simulated auditory representations are further corroborated by evaluating the overall signal-to-distortion fidelity (SI-SDR) of the neural responses, provided the appropriate loss functions are applied. When computing the SI-SDR metric directly on the NAP representations rather than the standard audio waveform, we find that standard MHA processing consistently distorts the neural signal more than the baseline condition. In fact, all tested MHA variants yielded worse NAP SI-SDR values. In contrast, SEANet configurations successfully achieved fidelity improvements in the NAP domain, but only when the \texttt{l1-nap} loss was utilized during training. Conversely, SEANet models optimized without this structural penalty (\texttt{pnd} and \texttt{ssim}) failed to improve upon the baseline as measured on the NAP, resulting in degraded SI-SDR values. 

Together, these results demonstrate that directly optimizing deep enhancement networks using auditory-domain losses provides a distinct advantage over traditional amplification strategies at restoring the overall fidelity and structural integrity of neural speech representations in noise.

\FloatBarrier

\section{Discussion, Conclusions, and Limitations}
\subsection{Discussion.} 

\textcolor{black!80}{\textbf{Overcoming the Limits of Conventional Architectures}}\\[1.5pt]
In brief, the work reported here demonstrates how the DAL provides a framework for shifting from the fixed architecture of conventional hearing aid signal processing to a more flexible, spectro-temporally dynamic ML-based approach. The incorporation of a sophisticated cochlea model enables a broad range of cochlear dysfunctions to be emulated, thus providing a target for compensation of the causes of hearing loss, rather than simply the symptomatic outcome of reduced sensitivity. Further, the choice of a lightweight, waveform-to-waveform DNN as a basis for the ML-based signal processing facilitates a more rapid migration to the highly resource-constrained hearing-aid platforms.

Importantly, the observed improvements are not attributable to the CARFAC-derived losses alone. The MHA baselines were also optimized under auditory-domain objectives, but their constrained filterbank and compression architecture limits the class of signal transformations they can implement. In contrast, SEANet provides a richer waveform-to-waveform mapping, allowing the model to learn nonlinear, context-dependent enhancement patterns that are difficult to express with conventional multi-band gain and compression. Thus, the gains reported here arise from the combination of biologically grounded optimization targets and a sufficiently expressive neural signal-processing architecture.

\textcolor{black!80}{\textbf{NAP- versus SAI-Domain Optimization}}\\[1.5pt]
Within these biologically grounded targets, the choice of optimization domain dictates how the network handles temporal fine structure. Training with a pointwise loss on NAPs, such as L1 distance, encourages the aided/impaired NAP to align coherently with the normal-hearing reference NAP. For this objective to be effective, the trainable hearing-aid algorithm must remain close to zero phase, or linear phase with delay compensation, so that temporal fine structure is aligned sample-by-sample. In contrast, training with a loss in the SAI domain relaxes this requirement: the aided/impaired NAP need not match the normal-hearing NAP pointwise in temporal fine structure, provided that it produces a near-normal SAI. This distinction explains the difference between objective measures that require temporal fine-structure alignment, such as NAP-domain L1 distance and Pearson correlation, and measures computed in the SAI domain. As observed in our results, NAP-domain metrics favor models trained with an L1 NAP loss, whereas SAI-domain metrics show that adding an L1 NAP term to an SAI-based loss yields comparatively small additional gains.

\textcolor{black!80}{\textbf{Correcting Cochlear Encoding Distortions}}\\[1.5pt]
Given the characterisation of hearing loss as encoding distortions in addition to reduced sensitivity \cite{Plomp1978-fd}, the fine spectro-temporal fidelity of the signal modifications enabled by the SEANet--CARFAC combination provide the potential to correct for these encoding distortions at the level of the cochlea -- a capability that is absent in current technology. Decreasing the distance between the NAPs from each model is a significant step towards reasserting the perceptual fidelity of speech and would represent a major advance, as current technology fails to provide adequate support for hearing impaired listeners in acoustically complex environments (the ``cocktail party'' problem), which is often when they need assistance most.

\textcolor{black!80}{\textbf{A Model-Agnostic Platform for Hyper-Personalization}}\\[1.5pt]
These experiments provide a proof of concept using only a relatively unsophisticated model of hearing dysfunction (limited to OHC functionality). More sophisticated models of cochlear dysfunction are enabled by other CARFAC parameters relating to different cochlear processes or even complex combinations of processes. This will allow ML-based hearing aids to be hyper-personalised to the needs of a specific listener, going well beyond the fittings using prescription based on the threshold audiogram. 

The DAL architecture is model agnostic. While this specific example utilized a low-latency SEANet model to satisfy real-time embedded constraints, the integrated CARFAC model can function as a universal differentiable trainer for any advanced neural-network framework. Likewise, the CARFAC cochlear model itself could also be substituted by any functional or phenomenological model that estimates the NAP output from the cochlea. This open-platform approach therefore promotes rapid iteration and exploration of many computational and machine-learning approaches for driving effective functional advances in hearing-assistance technologies. 

More broadly, DAL could represent a potentially transformative framework for hearing-aid design: rather than fitting devices only to compensate for audiometric symptoms, it enables hearing-aid algorithms to be optimized against individualized models of auditory neural encoding. This opens the possibility of hyper-personalized compensation for specific biological impairments, targeting the underlying cochlear mechanisms of hearing loss rather than only their downstream perceptual consequences.

\subsection{Conclusions.}  The DAL framework demonstrates how a computational model of hearing function (CARFAC) can be used to train a lightweight low latency CNN (SEANet) using gradient descent. The resulting biologically inspired training loop outperforms the tested MHA baselines in this simulated OHC-loss setting and improves simulated neural signal recovery when used to train a DNN for mild hearing impairment. This supports the use of a cochlear modeling approach to hearing-aid design and fitting for suprathreshold listening, as opposed to the conventional perceptually based models used to fit hearing aids using the threshold audiogram.

\subsection{Limitations.} The experiments described here provide a proof of concept using a relatively unsophisticated model of hearing dysfunction (limited to OHC functionality) which represents only a single type of mild, real-world hearing loss. The loss functions tested have yet been shown to correspond to real-world perceived improvements or improvement in objective measures of intelligibility and quality of speech in noise. These limitations will be addressed in future work, which will involve clinical A/B testing on hearing-impaired subjects. Future research must also ensure training datasets are sufficiently diverse to prevent algorithmic bias across different languages and accents.

An important consideration that we have not yet explored is the tradeoff around how much denoising is appropriate.  Noise suppression generally hurts intelligibility for normal-hearing listeners, because it can suppress weak low-SNR cues that normal-hearing listeners have access to.  For hearing-impaired users, on the other hand, noise suppression often improves speech intelligibility \cite{diehl2023restoring}.  This suggests that there is a ``sweet spot'' of how much to suppress noise, related to the degree of hearing impairment.  The user will likely want to be able to hear non-speech sounds as well, so limiting the amount of noise reduction may be indicated for that reason, too.  Using clean speech as the target is thus not necessarily the right ``ideal''.  Evaluating this tradeoff with human listeners might be a slow and expensive process, so simply giving users a preference-based noise-suppression control might be a solution.  

\section*{Code Availability}
The DAL framework, including dataset preparation, TFRecord generation, training and evaluation scripts, and reproducibility instructions, is available as part of the Australian Future Hearing Initiative \texttt{hp-acoustic} repository:
\url{https://github.com/Australian-Future-Hearing-Initiative/hp-acoustic/tree/main/Frameworks/DAL_framework}.

\section*{References}

% \bibliographystyle{unsrtnat}
% \bibliography{arxiv_2026}
% PASTE EVERYTHING HERE instead for arxiv submission:

%%%%%%%%%%%%%%%%%%%%%%%%%%%%%%%%%%%%%%%%%%%%%%%%%%%%%%%%%%%%

\appendix
%...

%%%%%%%%%%%%%%%%%%%%%%%%%%%%%%%%%%%%%%%%%%%%%%%%%%%%%%%%%%%%

\appendix

\section{Technical appendices and supplementary material}
\textbf{A. Distributed JAX/Flax Platform Implementation}\\[4pt]
Executing a continuous double-ended CARFAC neural mapping loop requires specialized hardware parallelization strategies to accommodate the massive sequential digital-filter state vectors. Because the filter cascade involves nonlinear and recursive digital updates across about 80 channels per audio frame, executing temporal gradient computations via standard unrolled automatic differentiation quickly exceeds GPU memory limits. To resolve this safely, our architecture utilizes a dedicated multi-device JAX platform distributing the continuous SEANet pre-training and the two-pass reference topology across separate TPU clusters via jax.pmap. State elements representing instantaneous basilar displacement filter state, complex pole radii, AGC feedback filter state, inner hair cell state, etc. are managed across isolated execution devices with continuous state carryover across sequential audio chunks. This ensures stable temporal fine-structure continuity without triggering out-of-memory gradient errors over long phonetic intervals.\\[6pt]

\textbf{B. Theoretical Derivations of Partial Normalization Distance}\\[4pt]
Let $\mathbf{X},\mathbf{Y}\in\mathbb{R}^{H\times W}$ denote two SAI frames, 
with row index $i=1,\ldots,H$ (frequency channels) and column index $j=1,\ldots,W$ (time samples). 
For any matrix $\mathbf{Z}$, define its global mean:
\begin{equation}
\bar{Z}=\frac{1}{HW}\sum_{i=1}^H\sum_{j=1}^W Z_{ij},
\end{equation}
and its row-wise mean:
\begin{equation}
\bar{Z}_i=\frac{1}{W}\sum_{j=1}^W Z_{ij},\qquad i=1,\ldots,H.
\end{equation}

Given exponents $\alpha,\beta\in[0,1]$ and stabilizers $\epsilon_a,\epsilon_b>0$, define the partially normalized entries:
\begin{equation}
\widehat{Z}_{ij} \;=\; 
\frac{Z_{ij}}{\big(\epsilon_a+\bar{Z}\big)^{\alpha}\,\big(\epsilon_b+\bar{Z}_i\big)^{\beta}}.
\end{equation}

The \emph{PN distance} between $\mathbf{X}$ and $\mathbf{Y}$ is defined as:
\begin{equation}
D_{\mathrm{PN}}(\mathbf{X},\mathbf{Y})
= \Big( \mathbb{E}\big[(\widehat{X}_{ij}-\widehat{Y}_{ij})^2\big] \Big)^{1/2},
\end{equation}
where the expectation $\mathbb{E}$ is taken over the empirical distribution of indices $(i,j)$. In particular, $(\alpha,\beta)=(0,0)$ yields the Euclidean distance, 
while $(\alpha,\beta)=(1,1)$ corresponds to full normalization.
\\[6pt]

% \section*{C. Mean performance of SEANet models versus baseline and MHA in the NAP domain}
% \begin{table}[H] % Note the capital H
%   \label{tab:nap-results}
%   \centering
%   \begin{tabular}{llll}
%   \toprule
%   Experiment & \begin{tabular}[b]{@{}l@{}}L1 Distance \\ (Lower=Better)\end{tabular} & ... \\
%   \midrule
%   Baseline (Noisy) & 0.099 & 0.428 & -6.009 \\[4pt]
  
%   mha (nl2) & 0.120 & 0.391 & -7.750 \\
%   mha (l1-nap) & 0.146 & 0.370 & -8.494 \\ 
%   mha (pnd) & 0.183 & 0.341 & -9.455 \\
%   mha (l1-nap + pnd) & 0.177 & 0.348 & -9.285 \\[4pt]

%   \bottomrule
%   \end{tabular}
% \end{table}

\section*{C. Mean performance of SEANet models versus baseline and MHA in the NAP domain}

\begin{table}[H]
  \centering
  \label{tab:nap-results}
  \begin{tabular}{lccc} 
  \toprule
  Experiment & 
  \makecell[b]{L1 Distance \\ (Lower=Better)} & 
  \makecell[b]{Pearson's Corr. \\ (Higher=Better)} & 
  \makecell[b]{SI-SDR \\ (Lower=Better)} \\
  \midrule
  Baseline (Noisy)   & 0.099 & 0.428 & -6.009 \\[4pt]
  
  mha (nl2)          & 0.120 & 0.391 & -7.750 \\
  mha (l1-nap)       & 0.146 & 0.370 & -8.494 \\ 
  mha (pnd)          & 0.183 & 0.341 & -9.455 \\
  mha (l1-nap + pnd) & 0.177 & 0.348 & -9.285 \\[4pt]

  seanet (l1-nap)  & \textbf{0.073} & \textbf{0.706} & -0.261 \\
  seanet (pnd) & 0.109 & 0.448 & -5.660 \\
  seanet (ssim) & 0.132 & 0.401 & -6.740 \\
  seanet (l1-nap + pnd) & 0.081 & 0.684 & -0.611 \\
  seanet (l1-nap + SSIM) & 0.085 & 0.690 & \textbf{-0.233} \\
  \bottomrule
  \end{tabular}
\end{table}

\section*{D. Mean performance of SEANet models versus baseline and MHA in the SAI domain}
\begin{table}[H] % Note the capital H
  \centering
  \label{tab:sai-results}
  \begin{tabular}{lccc} 
  \toprule
  Experiment & 
  \makecell[b]{L1 Distance \\ (Lower=Better)} & 
  \makecell[b]{Pearson's Corr. \\ (Higher=Better)}\\
  \midrule
  Baseline (Noisy) & 0.023 & 0.645 \\[4pt]
  
  mha (nl2) & 0.025 & 0.675\\
  mha (l1-nap) & 0.028 & 0.678\\ 
  mha (pnd) & 0.030 & 0.753\\
  mha (l1-nap + pnd) & 0.030 & 0.756\\[4pt]

  seanet (l1-nap)  & 0.017 & 0.853 \\
  seanet (pnd) & 0.016 & 0.864\\
  seanet (ssim) & \textbf{0.014} & \textbf{0.870}\\
  seanet (l1-nap + pnd) & 0.017 & 0.865\\
  seanet (l1-nap + SSIM) & \textbf{0.014} & 0.861 \\

  \bottomrule
  \end{tabular}
\end{table}

%%%%%%%%%%%%%%%%%%%%%%%%%%%%%%%%%%%%%%%%%%%%%%%%%%%%%%%%%%%%
\end{document}